\documentclass[fleqn,twoside]{article}
\usepackage{espcrc2}

\usepackage{graphicx}
\usepackage[figuresright]{rotating}

\newcommand{\AmS}{{\protect\the\textfont2
  A\kern-.1667em\lower.5ex\hbox{M}\kern-.125emS}}

\def\ls{{_<\atop^{\sim}}}
\def\gs{{_>\atop^{\sim}}}
\def\cgs{ ${\rm erg~cm}^{-2}~{\rm s}^{-1}$ }
\def\aap{A\&A}
\def\apj{ApJ}
\def\aj{AJ}
\def\apjs{ApJS}
\def\mnras{MNRAS}
\def\nat{Nature}

\title{High Energy Large Area Surveys: from BeppoSAX to Chandra and 
XMM-Newton}

\author{Fabrizio Fiore\address[OAR]{INAF - Osservatorio Astronomico di Roma\\
        via Frascati 33, Monteporzio (Rm) I00040 Italy}%
        \thanks{This work has been supported by ASI contracts I/R/107/00, 
	I/R/037/01, and by Cofin--99--034, and CNAA 2000, 2001 grants}}
       
\begin{document}

\begin{abstract}
Hard X-ray observations are the most efficient way to discriminate
accretion-powered sources from star-light. Furthermore, hard X-rays
are less affected than other bands by obscuration. For these reasons
the advent of imaging instruments above 2 keV, has permitted to
dramatically improve our understanding of accretion-powered sources
and their cosmic evolution. By minimizing the problems of AGN
selection and nuclear obscuration, the combination of deep and shallow
hard X-ray surveys, performed first with ASCA and BeppoSAX and then
with Chandra and XMM (e.g. ASCA LSS, HELLAS, CDFN, CDFS, Lockman Hole,
SSA13, HELLAS2XMM etc.), allows a detailed study of the evolution of
accreting sources.  Somewhat surprising results are emerging: 1) the
sources making the Cosmic X-ray Background peak at a redshift (z=0.7-1)
lower than soft X-ray selected sources and lower than predicted by
synthesis models for the CXB; 2) there is strong evidence of a
luminosity dependence of the evolution, low luminosity sources
(i.e. Seyfert galaxies) peaking at a significantly later cosmic time
than high luminosity sources.
\vspace{1pc}
\end{abstract}

\maketitle

\section{INTRODUCTION}

Hard X-ray surveys are the most direct probe of supermassive black
hole (SMBH) accretion activity, which is recorded in the Cosmic X-ray
Background (CXB), in wide ranges of SMBH masses, down to $\sim
10^6-10^7\,M_{\odot}$, and bolometric luminosities, down to $L\sim
10^{43}$ erg/s.  At z$\gs0.5$, these regimes of accretion are hardly
accessible to optical observations but may provide a significant
fraction of the total accretion power of the Universe.  X-ray surveys
can therefore be used to constrain the SMBH mass density,
models for the CXB \cite{setti89,coma95,coma01}, and models for
the formation and evolution of the structure in the universe
\cite{haehnelt03,menci03}.

The advent of imaging instruments in the 2-10 keV band, first aboard
ASCA and BeppoSAX and then on 
Chandra and XMM-Newton, has led to a
dramatic advance.  ASCA and BeppoSAX shallow surveys have resolved
about 20 \% of the 2-10 keV and 5-10 keV CXB, while Chandra and
XMM-Newton deep surveys have resolved 80-90\% of the 2-10 keV CXB
\cite{ueda99,aky00,ceca99,fiore99,fiore01,flf02,giommi00,mush00,giacc02,brandt01,hasi01,alex03}. A detailed study of
the cosmic evolution of the hard X-ray source population is being
pursued combining source identifications from both shallow and deep
surveys. These studies confirm, at least qualitatively, the
predictions of standard AGN synthesis models for the CXB: the 2-10 keV
CXB is mostly made by the superposition of obscured and unobscured AGN
\cite{fiore99,fiore03,flf02,barger01,barger02,hasi01,hasi03}.
Quantitatively, though, rather surprising results are emerging: a
rather narrow peak in the range z=0.7-1 is present in the redshift
distributions from ultra-deep Chandra and XMM-Newton pencil-beam
surveys, in contrast to the broader maximum observed in previous
shallower soft X-ray surveys (e.g. ROSAT, \cite{schmidt98,lehmann01})
and predicted by the above mentioned synthesis models; furthermore,
evidence is emerging (related to the difference above) of a luminosity
dependence in the number density evolution of both soft and hard X-ray
selected AGN.

The ultra-deep Chandra and XMM-Newton surveys of the Chandra Deep
Field North (CDFN \cite{brandt01}), Chandra Deep Field South (CDFS
\cite{giacc02}) and Lockman Hole (LH \cite{hasi01}) cover each
$\sim0.05-0.1$ deg$^2$. For this reason the number of high luminosity
sources in these surveys is small, being the slope of the AGN
luminosity function at high luminosities rather steep. As an example,
in the CDFN there are only 6 AGN with log(L$_{2-10keV}/erg s^{-1})\gs44$
at z$>3$ and 20 at z$>2$ \cite{cowie03}.  To compute an accurate
luminosity function on wide luminosity and redshift intervals, and to
find sizeable samples of ``rare'' objects, such as high luminosity,
highly obscured type 2 QSO or X-ray bright, optically normal galaxies,
XBONGs \cite{fiore00,coma02}, a much wider area needs to be covered,
of the order of a few deg$^2$.  To this purpose several high energy,
large area, medium-deep surveys are being pursued, like, for example,
the: HELLAS2XMM serendipitous survey, which, using XMM-Newton archival
observations \cite{baldi02} has the goal to cover $\sim4$ deg$^2$ at a
2-10 keV flux limit of a few$\times10^{-14}$ \cgs; the XMM Bright
Sample Survey, with the goal of covering $\sim50$ deg$^2$ at the flux
limit of a few $\times10^{-13}$ \cgs, using again XMM-Newton archive
observations; the ELAIS-S1 XMM survey, which covers a contiguous area
of 0.5 deg$^2$ at a flux limit of a few$\times10^{-15}$ \cgs; the
COSMOS-VIMOS-XMM survey, which will cover a contiguous area of $\sim2$
deg$^2$ at a similar flux limit.  Other relevant Chandra project
include: the Chandra serendipitous survey ``Champ'', which covers
$\sim14$ deg$^2$ at a 2-10 keV flux limit of a few$\times10^{-14}$\cgs
(\cite{kim03}); the Chandra serendipitous survey SEXSI
(\cite{harr03}); the Chandra-SWIRE survey of the Lockman field; the
extended Chandra survey of the CDFS region.

\section{THE BEPPOSAX HERITAGE}

BeppoSAX pioneered the study of the hard X-ray sky using
both its imaging telescopes (the 5-10 keV HELLAS survey) and
its collimated instruments (the 13-200 keV PDS survey of nearby AGN and
blank fields). We briefly review these two topics in the following.

\subsection{The hard X-ray sky faint source population}

\begin{figure}[htb]
\includegraphics[width=8cm]{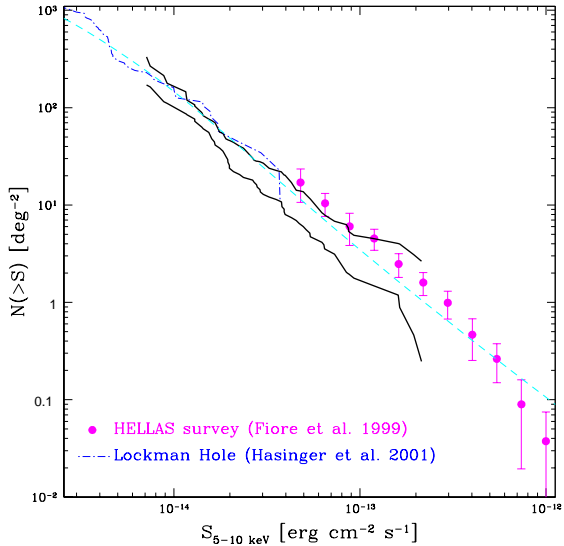}
\caption{The 5-10keV number counts from the BeppoSAX HELLAS survey,
filled points \cite{fiore01}, the HELLAS2XMM survey,
region inside the solid lines \cite{baldi02}, and the XMM
Lockman hole pointing, dot-dashed line \cite{hasi01}.}
\label{lnls}
\end{figure}

The BeppoSAX HELLAS survey covers $\sim85$ deg$^2$ of the sky down to
a 5-10 keV flux of $5\times10^{-14}$ \cgs \cite{fiore01}.
Fig. \ref{lnls} compares the integral 5-10 keV logN-logS of the 147
HELLAS sources with that obtained from deeper XMM surveys.  We have
obtained the optical identification of about half (62) of the sources
in a reduced area of sky (55 deg$^2$) with $\delta<+79^{o}$,
20hr$<\alpha<$5hr and 6.5hr$<\alpha<$17hr \cite{flf02}.  Because of
the quite large MECS error box ($\sim1'$ radius), we limited the
optical identification process to objects with surface density $\ls40$
deg$^{-2}$, to keep the number of spurious identifications in the
whole sample smaller than a few percent.  While broad line AGN are
identified up to z=2.76, all narrow line AGN have z$<$0.4, due to the
conservative threshold adopted for the optical magnitude of narrow
emission line AGN and galaxies (which do not show a bright optical
nucleus, because of the strong extinction).  The fraction of highly
obscured sources ($N_H\gs10^{23}$) at z$<0.3$ (where our survey should
be representative of the actual source population), is $\sim40\%$,
consistent with the expectations of AGN synthesis models
\cite{coma01}.  In this redshift range all highly obscured objects are
narrow line AGN and galaxies, while all broad line AGN have
$N_H<10^{23}$ cm$^{-2}$, consistent with popular AGN unification
schemes. The situation is somewhat different at high redshift, where
we found indications of low energy absorption in a few broad line AGNs
\cite{fiore01}.  Obscured, broad line AGNs at high z have been
discovered by ROSAT \cite{elvis94,fiore98} ASCA \cite{ceca99} and are
found in deeper XMM and Chandra surveys \cite{brusa03,page03}.
Possible explanations for the optical/X-ray dichotomy can be found in
\cite{maio01}.

\subsection{The BeppoSAX PDS view of highly obscured AGNs}

\begin{figure}[htb]
\includegraphics[height=8cm,angle=-90]{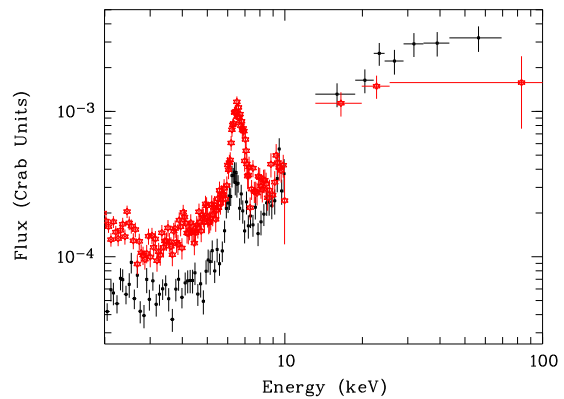}
\caption{Comparison of the BeppoSAX MECS and PDS spectra of
NGC1068 (open stars) with those of NGC6240 (dots). Both
spectra have been normalized to the Crab spectrum. Adapted
from Vignati et al. 1999 \cite{vig99}}
\label{n6240}
\end{figure}

The faint sources discovered by the BeppoSAX MECS are not accessible
to the BeppoSAX PDS \cite{fronte97}, which is limited to 13-80 keV
fluxes higher than $\sim10^{-11}$ \cgs. However, this is some 30 times
deeper than the HEAO1-A4 survey, and deeper than any previous (and
current!) satellite based experiment in this energy range. For this
reason the PDS observations of extragalactic sources have opened up a
new space of discovery and are a reference point for high energy AGN
studies.  The BeppoSAX PDS has detected some 100 AGN above 13 keV, a
factor of 5-10 improvement with respect to previous observations.
About 35 of these objects show strong obscuration at lower energies
\cite{matt00,risa02}, see also Matt, these proceedings.
Fig. \ref{n6240} show two examples, the archetypal Seyfert 2 galaxy
NGC1068 and the ultra-luminous infrared galaxy NGC6240. The presence
of a highly obscured, and intrinsically very luminous active nucleus
in the latter source was assessed thanks to this BeppoSAX PDS
observation \cite{vig99} (the optical spectrum of NGC6240 is typical
of a LINER).  Similar observations were performed by several other
authors, see e.g. \cite{matt00,fran99,maio98}.

The PDS has also performed a survey of about 200 deg$^2$ of the high
Galactic latitude sky down to its flux limit, using the off-source
pointings (the PDS consists of four phoswich units and is operated in
the so called ``rocking mode'', with a pair of units pointing to the
source while the other pair monitors the background $\pm210$ arcmin
away; the units on and off are interchanged every 96 seconds). A
fluctuation analysis of the PDS off-source pointings estimates in
0.02-0.06 the number of sources per deg$^2$ with 13-80 keV flux higher
than $10^{-11}$ \cgs. This is in agreement with the extrapolation of
the HELLAS 5-10 keV logN-logS in the 13-80 keV band if a power law
spectrum with $\alpha_E=0.8$ and reduced at low energy by a column
density of $1-3\times10^{23}$ cm$^{-2}$, consistent with the
expectation of AGN synthesis models for the CXB \cite{coma01}, is
assumed.

\section{THE HELLAS2XMM SURVEY}

The natural extension of the BeppoSAX HELLAS survey to $\sim10$ lower
X-ray fluxes, where the bulk of the CXB is resolved in sources, made
use of the higher throughput and much better source localization (a
few arcsec) of the XMM-Newton telescopes, which allowed the
identification of X-ray sources with faint optical counterparts.  We
have obtained optical photometric and spectroscopic follow-up of 122
sources in five XMM-Newton fields, covering a total of 0.9 deg$^2$
(the HELLAS2XMM `1dF' sample), down to a flux limit of
F$_{2-10keV}\sim10^{-14}$ \cgs.  We found optical counterparts
brighter than R$\sim25$ within $\sim6''$ from the X-ray position in
116 cases and obtained optical spectroscopic redshifts and
classification for 94 of these sources \cite{fiore03}.  The source
breakdown includes: 61 broad line QSO and Seyfert 1 galaxies; 14
narrow line AGN (9 of which have log(L$_{2-10keV}/erg s^{-1})>44$ and
can therefore be considered type 2 QSO); 14 emission line galaxies,
all with log(L$_{2-10keV}/erg s^{-1})>42.7$ and therefore all probably
hosting an AGN; 5 early type galaxies with $41.9<$log(L$_{2-10keV}/erg
s^{-1})<43.0$, therefore XBONGs, all probably hosting an AGN; 1 star; 2
groups or clusters of galaxies.  In the following we limit ourselves
to consider two broad categories: {\em optically unobscured AGN},
i.e. type 1, broad emission line AGN, and {\em optically obscured
AGN}, i.e. AGN whose nuclear optical emission, is totally or strongly
reduced by dust and gas in the nuclear region and/or in the host
galaxy.

\begin{figure}[h]
\includegraphics[height=8cm,angle=-90]{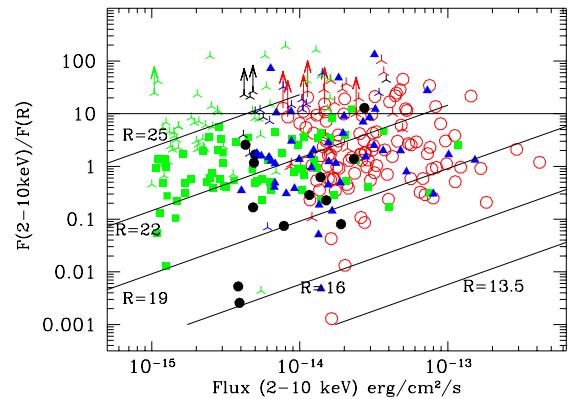}
\caption{The X-ray (2-10 keV) to optical (R band) flux ratio X/O as a
function of the X-ray flux for the combined sample (HELLAS2XMM = open
circles; CDFN = filled squares; LH = filled triangles; SSA13 = filled
circles, skeleton triangles are sources without a measured redshift).
Solid lines mark loci of constant R band magnitude.  The part of the
diagram below the R=25 line is accessible to optical spectroscopy with
10m class telescopes. Note that $\sim20\%$ of the sources have
X/O$\gs10$, irrespective of the X-ray flux.  HELLAS2XMM 1dF sources
with X/O$\gs10$ have R=24-25, and therefore their redshifts can be
measured through optical spectroscopy. }
\label{xos}
\end{figure}

We have combined the HELLAS2XMM 1dF sample with other deeper hard X-ray
samples including: 1- CDFN sample from \cite{barger02}: 120 sources
with flux F$_{2-10keV}>10^{-15}$ \cgs, 67 with a spectroscopic
redshift.  2- Lockman Hole sample from \cite{main02,baldi02}: 55
sources with F$_{2-10keV}>4\times10^{-15}$ \cgs, 41 with a
spectroscopic redshifts and 3 with a photometric redshifts; 3- SSA13
sample from \cite{mush00,barger01}: 20 sources with
F$_{2-10keV}>3.8\times10^{-15}$ \cgs, 13 with spectroscopic redshift.
Overall, we dealt with 317 faint hard X-ray selected sources, 221
(70\%) of them identified with an optical counterpart whose redshift
is available.  This ``combined'' sample includes 113 broad line AGN
and 108 optically obscured AGN.

Fig. \ref{xos} shows the X-ray (2-10 keV) to optical (R band) flux
ratio (X/O) for the combined sample.  About 20\% of the sources have
X/O$\gs10$, i.e ten times or more higher than the X/O typical of
optically selected AGN.  At the flux limit of the HELLAS2XMM 1dF
sample several sources with X/O$\gs10$ have optical magnitudes
R=24-25, bright enough for reliable spectroscopic redshifts to be
obtained with 8m class telescopes.  Indeed, we were able to obtain
spectroscopic redshifts and classification of 13 out of the 28
HELLAS2XMM 1dF sources with X/O$>10$: 8 of them are type 2 QSO at
z=0.7-1.8, to be compared with the total of 10 type 2 QSOs identified
in the CDFN \cite{cowie03} and CDFS \cite{hasi03}.

\begin{figure}[h]
\includegraphics[height=8cm,angle=-90]{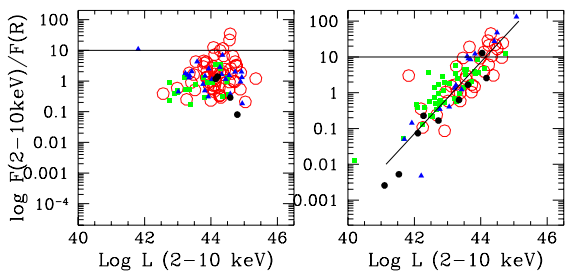}
\caption{The X-ray to optical flux ratio
versus the X-ray luminosity 
for type 1 AGN, left panel, and non type 1 AGN and galaxies,
right panel. Symbols as in Fig. \ref{xos}. The orizontal
lines mark the level of X/O=10, $\sim20\%$ of the sources in the
combined sample have X/O higher than this value. The diagonal line in the
right panel is the best log(X/O)--log(L$_{2-10keV}$) linear regression.}
\label{xolx}
\end{figure}

Fig. \ref{xolx} show the X-ray to optical flux ratio as a function of
the X-ray luminosity for broad line AGN (left panel) and non broad
line AGN and galaxies (right panel).  While the X/O of the broad line
AGNs is not correlated with the luminosity, a striking correlation
between log(X/O) and log(L$_{2-10keV}$) is present for the obscured AGN:
higher X-ray luminosity, optically obscured AGN tend to have
higher X/O.  The solid diagonal line in the panel represents the best
linear regression between log(X/O) and log(L$_{2-10keV}$) The nuclear
optical-UV light is completely blocked, or strongly reduced in these
objects, unlike the X-ray light. Indeed, the optical R band light of
these objects is dominated by the host galaxy and therefore, {\em X/O
is roughly a ratio between the nuclear X-ray flux and the host galaxy
starlight flux}.

\section{THE EVOLUTION OF HARD X-RAY SELECTED AGNS}

We can use the fractions of obscured to unobscured objects in the
combined sample and the correlations in Fig. \ref{xolx} to attribute,
in a purely statistical sense, a luminosity, and therefore a redshift,
to the sources in the combined sample without optical spectroscopic
identification.  The full combined sample (including both the objects
with measured redshift and the objects with the statistically
estimated redshift), complemented at high fluxes by 66 sources from
the HEAO1 A2 all sky survey \cite{grossan92} with
F$_{2-10keV}>2\times10^{-11}$ \cgs, has been used to compute the
evolution of the number density of hard X-ray selected sources, using
the standard 1/V$_{max}$ method \cite{schmidt68}.

\begin{figure}[h]
\includegraphics[height=8cm]{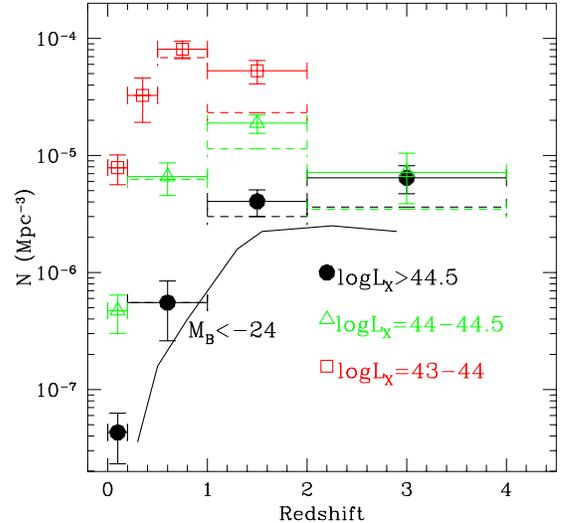}
\caption{The evolution of the number density of hard X-ray selected
sources in three bins of luminosity: log(L$_{2-10keV}/erg s^{-1})=43-44$ 
= empty squares; log(L$_{2-10keV}/erg s^{-1})=44-44.5$  =
empty triangles; log(L$_{2-10keV}/erg s^{-1})>44.5$ = filled circles.
Dashed lines represent lower limits obtained using only the sources
with measured redshift, see the text.  The solid continuous curve
represents the evolution of optically selected QSO more luminous than
M$_B=-24$ \cite{hs90}. Note that the shape of the solid
curve is similar to the evolution of the luminous X-ray selected
sources. }
\label{evol}
\end{figure}

Fig. \ref{evol} plots the evolution of the number density in three
luminosity bins: log(L$_{2-10 keV}/erg s^{-1})=43-44$, log(L$_{2-10
keV}/erg s^{-1})=44-44.5$ and log(L$_{2-10 keV}/erg s^{-1})=44.5-46$. The
dashed lines are lower limits computed using only the sources with
measured redshift.  We see that the number density of lower luminosity
AGN increases between z=0 and z=0.5 by a factor $\sim13$ and stays
approximately constant up to z$\sim2$.  Conversely, the number density
of luminous AGN increases by a factor $\sim100$ up to z=2 and by a
factor $\sim170$ up to z$\sim3$.  The last behaviour is similar to
that of luminous (M$_B<-24$) optically selected QSO \cite{hs90}.

\section{CONCLUSIONS}

Thanks to both deep, pencil beam, and shallow but larger area high
energy X-ray surveys performed in the past few years by BeppoSAX,
ASCA, Chandra and XMM-Newton we now have little doubts that the CXB is
due to the integrated contribution by (mainly) AGN and therefore that
it may be regarded as the electromagnetic outcome of mass accretion
onto SMBH in these galactic nuclei, along the cosmic history.  We have
been able to study the evolution of accreting sources of low
(i.e. Seyferts) and high luminosity (i.e. QSOs) up to z$\approx3$, to
find that there is strong evidence of a luminosity dependence of the
evolution, low luminosity sources (i.e. Seyfert galaxies) peaking at a
significantly later cosmic time than high luminosity sources (also see
\cite{hasi03} and Hasinger, these proceedings).  To push this study to
higher redshift and to derive separately the evolution of unobscured
and obscured objects we need to perform wide and deep surveys.  For
example, in the HELLAS2XMM 1dF sample there are some thirty "optically
obscured" AGN: 100-150 sources of this type would be sufficient to
adequately figure their luminosity function over 2-3 luminosity dex
and a few redshift bins. This is the goal of the extension of the
optical follow-up of the HELLAS2XMM survey from 1 to 4 deg$^2$,
which should allow us to contrast the luminosity function of obscured
and unobscured AGN, and to study their differential evolution up to
z$\sim$2.  On the other hand, the ultra-deep but small area CDFN
survey has provided so far only 6 AGN more luminous that
log(L$_{2-10keV}/erg s^{-1})=44$ at z$>3$ and 20 at z$>2$
\cite{cowie03}.  Note that all of them have flux $\gs10^{-15}$ \cgs,
suggesting that the most effective strategy to find high luminosity,
high z AGN consists in increasing the area covered at
F$_{2-10keV}=1-5\times10^{-15}$ \cgs, rather than pushing the depth of
the survey. Observing of the order of 1 deg$^2$ of sky at the above
flux limit, would roughly increase by a factor of 10 the number of
high z objects, a program which will be carried out by the ELAIS-S1
and COSMOS-VIMOS-XMM surveys.

The hard X-ray selected source's luminosity functions can be used to
determine the SMBH mass density. Using a simple approach
\cite{cowie03} we find a SMBH density of $4-5\times10^5$ M$_{\odot}$
Mpc$^{-3}$, about 2-3 times higher than that estimated using optical
surveys and by \cite{cowie03}, but consistent with both that measured
from the intensity of the CXB (\cite{fabian03} and references therein)
and from local galaxies \cite{gebh00,fm00}. This implies that most of
this accretion of mass takes place during `active' phases, when the
accreting matter radiates powerfully, giving rise to an AGN.  However,
below 10 keV we cannot see directly much ($\sim$50\%) of this
accretion luminosity, because it appears to be hidden behind layers of
gas and dust, Indeed, the energy range where most of the CXB energy
density resides (the 20-60 keV range) remains as of today essentially
unprobed, and therefore the fraction of highly obscured objects is
today poorly constrained.  The result is that all estimates on the
accretion luminosity in the Universe and of the SMBH mass density are
based on extrapolation of measurements performed below 10 keV into the
20-60 keV band, making rough assumptions about the fraction of
obscured AGNs. This situation stems from the lack, so far, of focusing
instruments in this band. A mission capable of exploring the hard
X-ray sky with focusing/imaging instrumentation would be able to reach
13-80 keV fluxes 100-200 times lower than those probed by the PDS, where
the source density is $\approx100$ sources deg$^{-2}$, corresponding
to resolving in sources roughly half of the CXB in that energy band.
This would therefore lead to the solution of the longest standing
issue in X-ray astronomy (and one of the most outstanding issue in
Cosmology), by making a great leap forward, comparable to that
achieved in the soft X-rays by the Einstein Observatory in the late
70'.

Acknowledgements. The original matter presented in this paper is the
result of the effort of a large number of people, in particular of the
HELLAS and HELLAS2XMM teams. I would like to thank the BeppoSAX SOC,
OCC and SDC teams for the successful operation of the BeppoSAX
satellite, preliminary data reduction and screaning, data calibration
and archiving.

\end{document}